\documentstyle[12pt]{article}
\def\Re{{\cal R \mskip-4mu \lower.1ex \hbox{\it e}\,}}
\def\Im{{\cal I \mskip-5mu \lower.1ex \hbox{\it m}\,}}

\def\beq{\begin{equation}}
\def\eeq{\end{equation}}

\def\sp6{Sp(6)_L \times U(1)_Y}

\def\Zbb{Z\rightarrow b\ov b}
\def\epsb{\epsilon_b}
\def\eps1{\epsilon_1}
\def\sub#1{_{\lower.25ex\hbox{$\scriptstyle#1$}}}
\def\sul#1{_{\kern-.1em#1}}
\def\sll#1{_{\kern-.2em#1}}
\def\sbl#1{_{\kern-.1em\lower.25ex\hbox{$\scriptstyle#1$}}}
\def\ssb#1{_{\lower.25ex\hbox{$\scriptscriptstyle#1$}}}
\def\sbb#1{_{\lower.4ex\hbox{$\scriptstyle#1$}}}

\def\MeV{\,{\rm MeV}}
\def\GeV{\,{\rm GeV}}

\def\JL{J. L. Lopez}
\def\DVN{D. V. Nanopoulos}

\def\to{\rightarrow}
\def\ov{\overline}
\def\mh{\ifmmode m\sbl H \else $m\sbl H$\fi}
\def\mch{\ifmmode m_{H^\pm} \else $m_{H^\pm}$\fi}
\def\mt{\ifmmode m_t\else $m_t$\fi}
\def\mc{\ifmmode m_c\else $m_c$\fi}
\def\mz{\ifmmode M_Z\else $M_Z$\fi}
\def\mw{\ifmmode M_W\else $M_W$\fi}
\def\mws{\ifmmode M_W^2 \else $M_W^2$\fi}
\def\mhs{\ifmmode m_H^2 \else $m_H^2$\fi}
\def\mzs{\ifmmode M_Z^2 \else $M_Z^2$\fi}
\def\mts{\ifmmode m_t^2 \else $m_t^2$\fi}
\def\mcs{\ifmmode m_c^2 \else $m_c^2$\fi}
\def\mchs{\ifmmode m_{H^\pm}^2 \else $m_{H^\pm}^2$\fi}
\def\ztwo{\ifmmode Z_2\else $Z_2$\fi}
\def\zone{\ifmmode Z_1\else $Z_1$\fi}
\def\mtwo{\ifmmode M_2\else $M_2$\fi}
\def\mone{\ifmmode M_1\else $M_1$\fi}
\def\tb{\ifmmode \tan\beta \else $\tan\beta$\fi}
\def\xw{\ifmmode x\sub w\else $x\sub w$\fi}
\def\ch{\ifmmode H^\pm \else $H^\pm$\fi}
\def\lum{\ifmmode {\cal L}\else ${\cal L}$\fi}
\def\inpb{\ifmmode {\rm pb}^{-1}\else ${\rm pb}^{-1}$\fi}
\def\infb{\ifmmode {\rm fb}^{-1}\else ${\rm fb}^{-1}$\fi}
\def\epem{\ifmmode e^+e^-\else $e^+e^-$\fi}
\def\ppb{\ifmmode \bar pp\else $\bar pp$\fi}

\newskip\zatskip \zatskip=0pt plus0pt minus0pt

\def\matth{\mathsurround=0pt}
\def\lsim{\mathrel{\mathpalette\atversim<}}
\def\gsim{\mathrel{\mathpalette\atversim>}}
\def\atversim#1#2{\lower0.7ex\vbox{\baselineskip\zatskip\lineskip\zatskip
  \lineskiplimit 0pt\ialign{$\matth#1\hfil##\hfil$\crcr#2\crcr\sim\crcr}}}

\renewcommand{\thefootnote}{\fnsymbol{footnote}}

\begin{document} \begin{titlepage}
\setcounter{page}{1}
\thispagestyle{empty}
\rightline{\vbox{\halign{&#\hfil\cr
&PURD-TH-94-08\cr
&SNUTP-94-46\cr
&May 1994\cr}}}
\begin{center}
\vglue 0.5cm
{\Large\bf Inclusion of $\Zbb$ vertex corrections  \\}
\vspace{0.2cm}
{\Large\bf in Precision Electroweak Tests \\}
\vspace{0.2cm}
{\Large\bf on the $Sp(6)_L \times U(1)_Y$ Model\\}
\vglue 1.0cm
{Gye~T.~Park$^{(a)}$ and T.~K.~Kuo$^{(b)}$\\}
\vglue 0.5cm
{\em $^{(a)}$Center for Theoretical Physics, Seoul National University\\}
{\em Seoul, 151-742, Korea\\}
{\em $^{(b)}$Department of Physics, Purdue University\\}
{\em West Lafayette, IN 47907, USA\\}
\baselineskip=12pt

\vglue 2.0cm
\end{center}

\begin{abstract}

We extend our previous work on the precision electroweak tests in the $Sp(6)_L
\times U(1)_Y$ family model to include for the first time the important
$\Zbb$ vertex corrections encoded in a new variable $\epsb$, utilizing all the
latest LEP data.
We include in our analysis the one loop EW radiative corrections due to the new
bosons in terms of $\epsilon_{1,b}$ and $\Delta\Gamma_Z$.
We find that the correlation between $\epsilon_1$ and $\epsb$
makes the combined constraint much stronger than the individual ones.
The model is consistent with the recent CDF result of $m_t=174\pm
10^{+13}_{-12}\GeV$, but it can not accomodate $m_t\gsim 195\GeV$.

\end{abstract}

\renewcommand{\thefootnote}{\arabic{footnote}} \end{titlepage}
\setcounter{page}{1}


\section{Introduction}

Having unknown top quark mass$(m_t)$ has long been one of the biggest
disadvantage in studying the Standard Model(SM) and its extensions.
With the very recent announcement of CDF collaboration at Tevatron on their
evidence for top quark production\cite{CDF-top} with $m_t=174\pm
10^{+13}_{-12}\GeV$, there are amusing possibilities that one can constrain
further possible new physics beyond the SM from the precision LEP data.
Precision measurements at the LEP have been remarkably successful in confirming
the validity of the SM\cite{1inUV}. Indeed, in order to have agreements between
theory and experiments, one has to go beyond the tree-level calculations and
include known electroweak(EW) radiative corrections. However, from the
theoretical point of view, there is a concensus that the SM can only be a low
energy limit of a more complete theory. It is thus of the utmost importance to
try and push to the limit in finding possible deviations from the SM. In fact,
there are systematic programs for such precision tests. Possible deviations
from the SM can all be summarized into a few parameters which then serve to
measure the effects of new physics beyond the SM. A lot of efforts have gone
into this type of investigation trying to develop a scheme to minimize the
disadvantage of having unkown $m_t$ but to optimize sensitivity to new physics.
To date significant constraints have been placed on a number of models, such as
the two Higgs
 the technicolor model\cite{RCTC}, and some extended gauge
models\cite{Altetal}.
In this work we wish to apply the analysis to another extension of the SM, the
$\sp6$ family model. Amongst several of the available parametrization schemes
in the literature, the most appropriate one for our purposes is that of
Altarelli et.~al\cite{ABJ}. This is because their $\epsilon$-parametrization
can be used for new physics which might appear at energy scales not far from
those of the SM. This is the case for the $\sp6$ model.
In fact, in an earlier analysis\cite{Kuo-park-eps}
we have performed precision EW tests in this model in a scheme using
$\epsilon_{1,2,3}$
introduced in Ref.~\cite{ABJ}.  We found that the parameters in this model were
severely constrained.
Recently, it was re-emphasized that there are important vertex corrections to
the decay mode $\Zbb$\cite{RC2HDM,ABCI,ABCII}. This mode has also been measured
at LEP and has proven to provide a strong constraint to model building.
Considering the high central value for the $m_t$ from CDF, $\Zbb$ vertex
corrections, which grow as $m_t^2$, can be quite significant.
Therefore, we have now incorporated $\Zbb$ vertex corrections in our new
analysis of the $\sp6$ model.
In this paper we extend our previous analysis in two aspects: (i) we include a
new parameter $\epsb$ to encode $\Zbb$ vertex corrections, (ii) we calculate
$\eps1$ in a new scheme introduced in Ref~\cite{ABCI} in order to take
advantage of all LEP data.
We find that inclusion of $\Zbb$ vertex corrections reinforces strongly the
previous constraint from $\eps1$ only so that the allowed parameter regions are
reduced considerably.
Thus, the precision EW tests have demonstrated clearly that they are powerful
tools in shaping our searches for
extensions of the SM.

In Sec.~II, we will describe the $\sp6$ model, spelling out in detail the parts
that are relevant to precision tests. In Sec.~III, we summarize properties of
the $\epsilon$-parameters which will be used in our analysis. Sec.~IV contains
our detailed numerical results. Finally, some concluding remarks are given in
Sec.~V.

\section{$\sp6$ Model}
The $SP(6)_L\otimes U(1)_Y$ model, proposed some time ago\cite{kuo-nakagawa},
is the simplest extension of the standard model of three generations
that unifies the standard $SU(2)_L$ with the horizontal gauge group
$G_H(=SU(3)_H)$ into an anomaly free, simple, Lie group. In this model,
the six left-handed quarks (or leptons) belong to a {\bf 6} of
$SP(6)_L$, while the right-handed fermions are all singlets. It is thus
a straightforward generalization of $SU(2)_L$ into $SP(6)_L$, with the three
doublets of $SU(2)_L$ coalescing into a sextet of $SP(6)_L$. Most of the
new gauge bosons are arranged to be heavy $(\geq 10^2$--$10^3\rm\,TeV)$ so as
to avoid sizable FCNC. $SP(6)_L$ can be naturally broken into $SU(2)_L$
through a chain of symmetry breakings. The breakdown
$SP(6)_L \rightarrow [SU(2)]^3 \rightarrow SU(2)_L$ can be induced by two
antisymmetric Higgs which tranform as $({\bf 1}, {\bf 14}, 0)$ under
$SU(3)_C\otimes SP(6)_L\otimes U(1)_Y$. The standard $SU(2)_L$ is to be
identified with the diagonal $SU(2)$ subgroup of
$[SU(2)]^3=SU(2)_1\otimes SU(2)_2\otimes SU(2)_3$, where $SU(2)_i$ operates
on the $i$th generation exclusively. In terms of the $SU(2)_i$ gauge boson
$\vec{A}_i$, the $SU(2)_L$ gauge bosons are given by $\vec{A}={1\over\sqrt 3}
(\vec{A}_1+\vec{A}_2+\vec{A}_3)$. Of the other orthogonal combinations of
$\vec{A}_i$,
$\vec{A}^\prime={1\over\sqrt 6}(\vec{A}_1+\vec{A}_2-2\vec{A}_3)$, which
exhibits unversality only
among the first two generations, can have a mass scale in the TeV range
\cite{1TeVZ}. The three gauge bosons $A^\prime$ will be denoted as $Z^\prime$
and $W^{\prime\pm}$.
Given these extra gauge bosons with mass in the TeV range, we can expect small
deviations from the SM. Some of these effects were already analyzed elsewhere.
For EW precision tests,
the dominant effects of new heavier gauge boson $Z^\prime (W^{\prime\pm})$ show
up
in its mixing with the standard $Z(W^\pm)$ to form the mass eigenstates
$Z_{1,2} (W_{1,2})$:
\[ \hbox to \hsize{$ \hfill
\begin{array}{rcl}
Z_1&=&Z\cos\phi_Z+Z^\prime\sin\phi_Z \;, \\
W_1&=&W\cos\phi_W+W^\prime\sin\phi_W \;,
\end{array} \quad
\begin{array}{rcl}
Z_2 &=& -Z\sin\phi_Z+Z^\prime\cos\phi_Z \;, \\
W_2 &=& -W\sin\phi_W+W^\prime\cos\phi_W \;,
\end{array} \hfill
\begin{array}{r}
\stepcounter{equation}(\theequation)\\
\stepcounter{equation}(\theequation)
\end{array}
$} \]
where $Z_1 (W_1)$ is identified with the physical $Z(W)$.
Here, the mixing angles $\phi_Z$ and $\phi_W$ are expected to be small
$(\lsim0.01)$, assuming that they scale as some powers of mass ratios.

With the additional gauge boson $Z^\prime$, the neutral-current Lagrangian
is generalized to contain an additional term
\begin{equation}
L_{NC}=g_Z J_Z^\mu Z_\mu +g_{Z^\prime} J_{Z^\prime}^\mu Z_\mu^\prime \;,
\end{equation}
where $g_{Z^\prime}=\sqrt{1-x_W\over 2} g_Z={g\over\sqrt{2}}$,
$x_W=\sin^2\theta _W$, and $g={e\over {\sin\theta _W}}$. The neutral currents
$J_Z$ and $J_{Z^\prime}$ are given by
\begin{eqnarray}
J_Z^\mu &=&\sum_{f} \bar{\psi}_f\gamma^\mu\left( g^f_V+g^f_A\gamma _5\right)
\psi_f \;, \\
J_{Z^\prime}^\mu &=&\sum_{f} \bar{\psi}_f\gamma^\mu\left( g^{\prime
f}_V+g^{\prime f}_A\gamma _5\right)
\psi_f \;,
\end{eqnarray}
where $g^f_V={1\over 2}\left( I_{3L}-2x_Wq\right)_f$, $g^f_A={1\over
2}\left( I_{3L}\right)_f$ as in SM, $g^{\prime f}_V=g^{\prime
f}_A={1\over 2}\left( I_{3L}\right)_f$
for the first two generations and $g^{\prime f}_V=g^{\prime
f}_A=-\left( I_{3L}\right)_f$ for the third. Here $\left(I_{3L}\right)_f$ and
$q_f$
are the third component of weak isospin and electric charge of fermion $f$,
respectively. And the neutral-current Lagrangian reads in terms of $Z_{1,2}$
\begin{equation}
L_{NC}=g_Z\sum_{i=1}^2\sum_{f} \bar{\psi}_f\gamma_\mu\left(
g^f_{Vi}+g^f_{Ai}\gamma _5\right)
\psi_f Z^\mu_i \;,
\end{equation}
where $g^f_{Vi}$ and $g^f_{Ai}$ are the vector and axial-vector
couplings of fermion $f$ to physical gauge boson $Z_i$, respectively.
They are given by
\begin{eqnarray}
g^f_{V1, A1}&=&g^f_{V, A}\cos\phi_Z+{g_{Z^\prime}\over g_Z} g^{\prime
f}_{V, A}\sin\phi_Z \;, \\
g^f_{V2, A2}&=&-g^f_{V, A}\sin\phi_Z+{g_{Z^\prime}\over g_Z} g^{\prime
f}_{V, A}\cos\phi_Z \;.
\end{eqnarray}
Similar analysis can be carried out in the charged sector.

\section{One-loop EW radiative corrections and the $\epsilon$-
parameters}
There are several different schemes to parametrize
the EW vacuum polarization corrections \cite{Kennedy,PT,efflagr,AB}. It
can be easily shown that by expanding the vacuum polarization tensors to order
$q^2$, one obtains three independent physical parameters. Alternatively, one
can
show that upon symmetry breaking there are three
additional terms in the effective lagrangian \cite{efflagr}.
In the $(S,T,U)$ scheme \cite{PT},
the deviations of the model predictions from those of the SM (with fixed
values of $m_t,m_H$) are considered to be  as the effects from ``new physics".
This scheme is
only valid to the lowest order in $q^2$, and is therefore not viable for a
theory with new, light $(\sim M_Z)$ particles. In the $\epsilon$-scheme, on the
other hand, the model predictions are absolute and are valid up to higher
orders in $q^2$, and therefore this scheme is better suited to
the EW precision tests of the MSSM\cite{BFC} and a class of supergravity models
\cite{PARKeps}.
Here we choose to use the $\epsilon$-scheme because the new particles in the
model to be considered here can be relatively light $(O(1TeV))$.

There are two different $\epsilon$-schemes. The original scheme\cite{ABJ} was
considered in our previous analysis \cite{Kuo-park-eps}, where
$\epsilon_{1,2,3}$ are defined from a basic set of observables $\Gamma_{l},
A^{l}_{FB}$ and $M_W/M_Z$.
Due to the large $m_t$-dependent vertex corrections to $\Gamma_b$, the
$\epsilon_{1,2,3}$ parameters   and $\Gamma_b$ can be correlated only for a
fixed value of $m_t$. Therefore, $\Gamma_{tot}$, $\Gamma_{hadron}$ and
$\Gamma_b$ were not included  in Ref.~\cite{ABJ}. However, in the new
$\epsilon$-scheme, introduced recently in Ref.~\cite{ABCI}, the above
difficulties are overcome by introducing a new parameter $\epsilon_b$ to encode
the $\Zbb$ vertex corrections. The four $\epsilon$'s are now defined from an
enlarged set of $\Gamma_{l}$, $\Gamma_{b}$, $A^{l}_{FB}$ and $M_W/M_Z$ without
even specifying $m_t$.
In this work we use this new $\epsilon$-scheme.
Experimental values for $\epsilon_1$ and $\epsilon_b$ are determined by
including all the latest LEP data(complete '92 LEP data+ preliminary '93 LEP
data) to be \cite{Altarelli}
\beq
\epsilon^{exp}_1=(-0.3\pm3.2)\times10^{-3},\qquad
\epsilon^{exp}_b=(3.1\pm5.5)\times10^{-3}\ .
\eeq

The expression for $\epsilon_1$ is given as
\cite{BFC}
\beq
\epsilon_1=e_1-e_5-{\delta G_{V,B}\over G}-4\delta g_A,\label{eps1}
\eeq
where $e_{1,5}$ are the following combinations of vacuum polarization
amplitudes
\begin{eqnarray}
e_1&=&{\alpha\over 4\pi \sin^2\theta_W M^2_W}[\Pi^{33}_T(0)-\Pi^{11}_T(0)],
\label{e1}\\
e_5&=& M_Z^2F^\prime_{ZZ}(M_Z^2),\label{e5}
\end{eqnarray}
and the $q^2\not=0$ contributions $F_{ij}(q^2)$ are defined by
\beq
\Pi^{ij}_T(q^2)=\Pi^{ij}_T(0)+q^2F_{ij}(q^2).
\eeq
The quantities $\delta g_{V,A}$ are the contributions to the vector and
axial-vector form factors at $q^2=M^2_Z$ in the $Z\to l^+l^-$ vertex from
proper vertex diagrams and fermion self-energies, and $\delta G_{V,B}$ comes
from the one-loop box, vertex and fermion self-energy corrections
to the $\mu$-decay amplitude at zero external momentum. It is important to note
that these non-oblique corrections are non-negligible.
Also, they must be included since in general only the physical observables
$\epsilon_i$, but not the individual terms in them, are
gauge-invariant\cite{gaugeinv}.
However, we have included the Standard non-oblique corrections only.
The contributions from the new physics are small, at least in the gauge that we
choose, and will be neglected here.

Following Ref.~\cite{ABCI}, $\Zbb$ vertex corrections are encoded in a new
variable $\epsb$ defined from $\Gamma_b$, the inclusive
partial width for $\Zbb$, as follows

\begin{equation}
\Gamma_b=3 R_{QCD} {G_FM^3_Z\over 6\pi\sqrt 2}\left(
1+{\alpha\over 12\pi}\right)\left[ \beta _b{\left( 3-\beta
^2_b\right)\over 2}(g^b_V)^2+\beta^3_b (g^b_A)^2\right] \;,
\end{equation}
with
\begin{eqnarray}
R_{QCD} &\cong&\left[1+{\alpha_S\left(
M_Z\right)\over\pi}-1.1{\left(\alpha_S\left(
M_Z\right)\over\pi\right)}^2-12.8{\left(\alpha_S\left(
M_Z\right)\over\pi\right)}^3\right] \;,\\
\beta_b&=&\sqrt {1-{4m_b^2\over M_Z^2}} \;, \\
g^b_A&=&-{1\over2}\left(1+{\epsilon_1\over2}\right)\left(
1+{\epsb}\right)\;,\\
{g^b_V\over{g^b_A}}&=&{{1-{4\over3}{\ov s}^2_W+\epsb}\over{1+\epsb}}\;.
\end{eqnarray}
Here ${\ov s}^2_W$ is an effective $\sin^2\theta_W$ for on-shell $Z$, and
$\epsb$ is closely related to the real part of the vertex correction to $\Zbb$,
denoted by $\delta_{b-Vertex}$ and defined explicitly in
Ref.~\cite{Bernabeuetal}.  In the SM, the diagrams for $\delta_{b-Vertex}$
involve top quarks and
$W^\pm$ bosons, and the contribution to $\epsb$ depends
quadratically on $m_t$, which is due to the EW symmetry breaking and can be a
decisive test of the model. In supersymmetric models there are additional
diagrams
involving Higgs bosons and supersymmetric particles\cite{BF,Djouadietal}.
In fact, $\epsilon_b$ has been calculated by one of us(G.P) in the context of
non-supersymmetric two Higgs doublet model\cite{epsb2HD}.

In the following section we
calculate $\epsilon_{1}$ and $\epsilon_{b}$ in the
$\sp6$ model. We do not, however, include $\epsilon_{2,3}$ in our analysis
simply because these parameters can not provide any constraints at the current
level
of experimental accuracy\cite{PARKeps}.
Although the oblique corrections due to extra gauge bosons could be neglected
completely as in Ref\cite{Altetal}, we have improved the model predictions for
the oblique corrections by implementing the new vertices from Eq.~(6)
for the fermion loops only.
In this way we have accounted for a significant deviation of the model
prediction from the SM value for not so small $|\phi_{Z,W}|$.
Furthermore, in models with extra gauge bosons such as the model to be
considered here, the contribution from the mixings of these extra bosons with
the SM ones $(\Delta\rho_M)$ should also be added to
$\epsilon_1$\cite{Altetal,Altarelli90,parkkuo93}.

\section{Results and Discussion}

In order to calculate the model prediction for the Z width, it is sufficient
for our purposes to resort to the improved Born approximation (IBA)\cite{IBA},
neglecting small additional effects from the new physics.
Weak corrections can be effectively included
within the IBA, wherein the
vector couplings of all the fermions are determined by an effective
weak mixing angle.
In the case $f\not= b$, vertex corrections are negligible, and one
obtains the standard partial $Z$ width
\begin{equation}
\Gamma(Z\longrightarrow f\bar{f})=N^f_C\rho {G_FM^3_Z\over 6\pi\sqrt 2}\left(
1+{3\alpha\over 4\pi}q^2_f\right)\left[ \beta _f{\left( 3-\beta
^2_f\right)\over 2}{g^f_{V1}}^2+\beta^3_f {g^f_{A1}}^2\right] \;,
\end{equation}
where $N_C^f =1$ for leptons, and for quarks
\begin{eqnarray}
N_C^f &\cong&3\left[1+{\alpha_S\left(
M_Z\right)\over\pi}-1.1{\left(\alpha_S\left(
M_Z\right)\over\pi\right)}^2-12.8{\left(\alpha_S\left(
M_Z\right)\over\pi\right)}^3\right] \;,\\
\beta_f&=&\sqrt {1-{4m_f^2\over M_Z^2}} \;, \\
\rho&=&1+\Delta\rho_M+\Delta\rho_{SB}+\Delta\rho_t\;, \\
\Delta\rho_t &\simeq& {3G_Fm_t^2\over 8\pi^2\sqrt 2} \;.
\end{eqnarray}
where the $\rho$ parameter includes not only the effects of the
symmetry breaking $\left(\Delta\rho_{SB}\right)$\cite{SB} and those
of the mixings between the SM bosons and the new bosons
$\left(\Delta\rho_{M}\right)$, but also the loop effects
$\left(\Delta\rho_{t}\right)$.
$N_C^f$ above is obtained by accounting for QCD corrections up to
3-loop order in $\overline{MS}$ scheme, and we ignore different QCD
corrections for vector and axial-vector couplings which are due not
only to chiral invariance broken by masses but also the large mass
splitting between $b$ and $t$. We use for the vector and axial vector couplings
$g^f_{V1}$ and $g^f_{A1}$ in Eq.~(7) the effective $\sin^2\theta_W$,
$\bar{x}_W=1-{M_W^2\over{\rho M_Z^2}}$.
In the case of $Z\longrightarrow b\bar{b}$,
the large $t$ vertex correction should be accounted for by the following
replacement
\begin{equation}
\rho \longrightarrow\rho-{4\over 3}\Delta\rho_t\,, \quad
\bar{x}_W\longrightarrow\bar{x}_W\left( 1+{2\over 3}\Delta\rho_t\right) \;.
\end{equation}

In the following analysis, we consider not only a constraint on the deviation
of $\Gamma_Z$ from the SM prediction\cite{parkkuo93}, $\Delta\Gamma_Z\leq
14$~MeV, which is the present experimental accuracy\cite{Luth}, but also the
present experimental
bound on $\Delta\rho_M$.
We use a direct model-independent bound on $\Delta\rho_M$,
$\Delta\rho_M\lsim 0.0147-0.0043{\left({m_t\over 120 GeV}\right)}^2$ from
$1-({M_W\over{M_Z}})^2=0.2257\pm 0.0017$ and $M_Z=91.187\pm
0.007\GeV$\cite{Luth}.
The values $M_H=100\GeV$, $\alpha_S(M_Z)=0.118$, and $\alpha(M_Z)=1/128.87$
will be used
thoughout the numerical analysis.

In Fig.~1 we present the model predictions for $\epsilon_1$ and $\epsilon_b$
only for the values of $\phi_Z$ and $\phi_W$ allowed by $\Delta\Gamma_Z$ and
$\Delta\rho_M$ constraints with $M_{Z^\prime}=1000$ and $M_{W^\prime}=800\GeV$
for $m_t=160, 175$ and  $190\GeV$. We restrict  $|\phi_{Z,W}|\leq 0.02$. We
also include in the figure the latest $90\%$CL ellipse from all LEP
data\cite{Altarelli}. The values of $m_t$ used are as indicated over each
horizontal stripe of dotts.
It is very interesting for one to see that the correlated contraint is much
stronger than individual constraints.
The maximum deviation of $\epsilon_b$ in the model from the SM value is around
$1.3\%$ for $|\phi_{Z,W}|\lsim0.02$. Although the deviation is very small, the
inclusion
of the $\epsilon_b$ in the analysis makes the LEP data certainly much more
constraining.
Imposing the $\epsilon_1-\epsilon_b$ constraint by selecting only the values of
$\phi_Z$ and $\phi_W$ falling inside the ellipse in Fig.~1, we show in Fig.~2
the allowed regions in $(\phi_Z, \phi_W)$ for (a) $m_t=160\GeV$, (b)
$m_t=175\GeV$ and (c) $m_t=190\GeV$. The striking difference in the shape of
the allowed region between $m_t=175\GeV$ and $m_t=190\GeV$ comes from the fact
that the SM value
of $\epsilon_1$ for $m_t=175\GeV$ is inside the ellipse whereas the one for
$m_t=190\GeV$ is outside\cite{Kuo-park-eps}.

If the top quark turns out to be fairly heavy, e.g. $m_t\gsim 180\GeV$
where the SM predictions always fall outside the ellipse in the Fig.~1
, then the presence of the extra gauge bosons is certainly favored
because it can bring the model predictions inside the ellipse
as seen in Fig.~1
 although there is an ambiguity in the model prediction for $\epsilon_1$ that
the contribution from the extra gauge bosons can have either signs. This
situation can be contrasted with the one in
the MSSM where the heavy top quark is still consistent with the LEP data as
long as the chargino
is very light $\sim M_Z/2$, which is known as ``light chargino
effect"\cite{BFC}, whose contribution to $\epsilon_1$ is always negative.
However, if the chargino were not discovered at LEP~II,
then MSSM would fall into a serious trouble.

\section{Conclusions}

In this work we have extended our previous work on the  precision EW tests in
the $\sp6$ family model to include for the first time the important
$\Zbb$ vertex corrections encoded in a new variable $\epsb$, utilizing all the
latest LEP data.
As has been the case with similar studies, the model is considerably
constrained. The most important effects of the model come from mixings of the
SM gauge bosons $Z$ and $W$ with the additional gauge bosons $Z^\prime$ and
$W^\prime$. We have included in our analysis the one loop EW radiative
corrections due to the new bosons in terms of $\epsilon_{1,b}$ and
$\Delta\Gamma_Z$.
It is found that the correlation between $\epsilon_1$ and $\epsb$
makes the combined $\epsilon_1-\epsb$ constraint much stronger than the
individual ones.
Using a global
fit to LEP data on $\Gamma_{l}, \Gamma_{b}, A^{l}_{FB}$ and $M_W/M_Z$
measurement, we find that the mixing angles $\phi_Z$ and $\phi_W$ are
constrained to lie in rather small regions. Also, larger ($\gsim 1\%$)
$\phi_Z$ and $\phi_W$ values are allowed only when there is considerable
cancellation between the $Z^\prime$ and $W^\prime$ contributions, corresponding
to $|\phi_Z|\approx|\phi_W|$. It is noteworthy that the results are sensitive
to the top quark mass.
For smaller $m_t$'s, the allowed parameter regions become considerably larger.
Only very tiny regions are allowed for $m_t=190\GeV$.
It is very interesting for one to see that the model can not accomodate
$m_t\gsim 195\GeV$ at $90\%$CL, which is still consistent with the $m_t$ from
CDF.
As the top quark mass from the Tevatron becomes more accurate, we can narrow
down the mixing angles further with considerable precision.

\section*{Acknowledgements}

The authors would like to thank professor J.~E.~Kim
for reading the manuscript.
The work of T.~K. has been supported in part by DOE. The work of G.~P.
has been supported by the Korea Science and Engineering Foundation
through the SRC program.

\newpage

%
\def\NPB#1#2#3{Nucl. Phys. B {\bf#1} (19#2) #3}
\def\PLB#1#2#3{Phys. Lett. B {\bf#1} (19#2) #3}
\def\PLIBID#1#2#3{B {\bf#1} (19#2) #3}
\def\PRD#1#2#3{Phys. Rev. D {\bf#1} (19#2) #3}
\def\PRL#1#2#3{Phys. Rev. Lett. {\bf#1} (19#2) #3}
\def\PRT#1#2#3{Phys. Rep. {\bf#1} (19#2) #3}
\def\MODA#1#2#3{Mod. Phys. Lett. A {\bf#1} (19#2) #3}
\def\IJMP#1#2#3{Int. J. Mod. Phys. A {\bf#1} (19#2) #3}
\def\TAMU#1{Texas A \& M University preprint CTP-TAMU-#1}
\def\PURD#1{Purdue University preprint PURD-TH-#1}
\def\ARAA#1#2#3{Ann. Rev. Astron. Astrophys. {\bf#1} (19#2) #3}
\def\ARNP#1#2#3{Ann. Rev. Nucl. Part. Sci. {\bf#1} (19#2) #3}

\newpage

%
{\bf Figure Captions}
\begin{itemize}

\item Figure 1: The correlated predictions for the $\epsilon_1$ and
$\epsilon_b$ parameters in the unit of $10^{-3}$ in the $Sp(6)_L\times U(1)_Y$
model.
The ellipse represents the $90\%$CL contour obtained from all LEP data
including the preliminary 1993 data. The values of $m_t$ are as indicated.
$M_{Z^\prime}=1000\GeV$ and $M_{W^\prime}=800\GeV$ are used. The dotts
represent the values of $\phi_Z$ and $\phi_W$  allowed by
$\Delta\Gamma_Z\leq 14 \MeV$ and $\Delta\rho_M$ constraint with
$|\phi_{Z,W}|\leq 0.02$.
\item Figure 2: The model parameter space allowed by
the further contraint from $\epsilon_1$-$\epsilon_b$ using
the $90\%$CL contour
in Figure 1 for (a) $m_t=160\GeV$, (b) $m_t=175\GeV$ and (c) $m_t=190\GeV$.
$M_{Z^\prime}=1000\GeV$ and $M_{W^\prime}=800\GeV$ are used.
\end{itemize}


\begin{thebibliography}{99}
\bibitem{SB} We assume that $\rho_{SB}=1$ even if it depends on a
particular Higgs structure.
\end{thebibliography}
\end{document}